\documentclass[]{aastex631}
\usepackage{graphicx}%
\graphicspath{{./}{./}}
\usepackage{color}
\usepackage{amsmath}  %
\usepackage{verbatim}     %
\usepackage[noabbrev,capitalise]{cleveref}  %
\usepackage{natbib}
\usepackage[mathscr]{euscript} %
\usepackage{xspace}
\usepackage{tabularx}
\makeatletter
\usepackage{etoolbox}
\patchcmd\H@refstepcounter{\protected@edef}{\protected@xdef}{}{}
\makeatother

\newcommand{\plasticc}{\textit{PLAsTiCC}\xspace}

\newcommand{\h}{\ensuremath{\text{H}_0}\xspace}

\newcommand{\w}{\ensuremath{w_0}\xspace}

\newcommand{\romanST}{\textit{Roman Space Telescope}\xspace}
\newcommand{\rst}{\textit{Roman}\xspace}
\newcommand{\romanplasticc}{Hourglass\xspace}

\newcommand{\Parsnip}{ParSNIP\xspace}

\usepackage{float}
\usepackage{cleveref}
\usepackage{placeins} %

\usepackage{hyperref} %

\crefname{figure}{Figure}{Figures}
\Crefname{figure}{Figure}{Figures}

\begin{document}

\title{Picture Perfect: Photometric Transient Classification Using the \Parsnip Model with \rst \romanplasticc Simulations}

\author[0009-0001-9439-4789]{Belal Abdelhadi}
\affiliation{Department of Physics and Astronomy, University of Hawai`i at M$\bar{a}$noa, Honolulu, Hawai`i 96822, USA}

\author{David Rubin}
\affiliation{Department of Physics and Astronomy, University of Hawai`i at M$\bar{a}$noa, Honolulu, Hawai`i 96822, USA}

\begin{abstract}

The \romanST, equipped with a 2.4-meter primary mirror and optical--NIR wide-field camera, promises to revolutionize our understanding of dark energy, exoplanets, and infrared astrophysics. One of the \rst Core Community Surveys is the High Latitude Time Domain Survey (HLTDS), which will measure more than 10,000 SN~Ia light curves but obtain a fraction of this number with spectra. The remaining SNe will have to be photometrically classified to achieve the full potential of the \rst HLTDS. To investigate transient yields and classifications, \citet{Rose2023} updated the Photometric LSST Astronomical Time-series Classification Challenge (\plasticc) framework (originally developed for the Vera Rubin Observatory) for the \rst HLTDS. This study leverages this \rst ``\romanplasticc'' dataset to train and evaluate the \Parsnip (Parameterized Supernova Identification Pipeline) model. We employ this model to classify various transient types from photometric data, paying particular attention to the types most represented in the dataset: normal SNe~Ia, 91bg-like SNe~Ia, SNe~Iax, and CC~SNe. The \Parsnip model's performance is assessed through confusion matrices and ROC curves across different redshift ranges. Our analysis reveals that while the model performs robustly at higher redshifts (with the AUC for classification varying between 0.9 and 0.95 in the range $0.5 \lesssim z \lesssim 2$), its accuracy dips at the lowest redshifts of the survey, likely due to limited training data. These findings underscore the importance of ensuring adequate representation of classes in the training set. This work underscores the value of machine learning models for next-generation surveys, paving the way for future studies with the \romanST for survey optimization, cosmological forecasts, and synergies with other surveys.
\end{abstract}


\section{Introduction} \label{sec:intro}

The \romanST, with its large 2.4-meter primary mirror and wide-field camera \citep{Spergel2015}, represents a significant advancement in astronomical surveys. Covering 100 times the area of the Hubble Space Telescope in a single exposure while maintaining similar resolution, and with much lower overheads, \rst is set to explore dark energy, exoplanets, and infrared astrophysics. This capability for large-scale time-domain surveys makes it a powerful tool for detecting transient astronomical events, promising unprecedented insights into the dynamic universe.

Three Core Community Surveys are currently planned: the High Latitude Time Domain Survey (HLTDS) which is the subject of this work, the High Latitude Wide Area Survey focused on static cosmological science, and the Galactic Bulge Time Domain Survey Survey focused on discovering exoplanets through microlensing. (There is also a Galactic Plane General Astrophysics Survey being defined through a similar community process.) The HLTDS will collect more than 10,000 type Ia supernova (SN~Ia) light curves but will only obtain spectra for a fraction (e.g., \citealt{Rose2021c}). For the remaining events, accurate photometric classification will be essential to fully utilize the survey's potential. 

Photometric classification, particularly at higher redshifts, is difficult due to observational biases in the training sets, sparsely sampled (or incomplete) light curves, and a lack of signal to noise. To tackle challenges like this, machine learning models are increasingly being used to photometrically classify transient astronomical events and many different approaches are being investigated. SNe Ia are the most developed, with several different empirical models that accurately describe light curves and spectra \citep[e.g.,][]{Guy2007, Saunders2018,Leget2020, Mandel2022}, however similar models for other transients are lacking. Some studies have explored feature extraction methods to distinguish light curves \citep{Bazin2009, Sanders2015, Guillochon2018, Lochner2016, Muthukrishna2019, Moller2020}. However, these methods are highly dependent on observing conditions, especially redshift, often leading to datasets biased toward brighter, lower-redshift objects, making photometric classification particularly challenging \citep{Lochner2016, Boone2019}. Machine learning has already proven its value in astronomy through several key studies. For example, \citet{Villar2021} applied deep learning for live anomaly detection of extragalactic transients, demonstrating how machine learning can identify rare events. Other recent studies, such as those by \citet{Aleo2023} and \citet{Alves2023}, have also focused on early supernova detection and the impact of observational cadence. These works highlight important strategies for maximizing transient detection efficiency, a critical component of time-domain surveys. \citet{Qu2021} and \citet{Pimentel2023} use neural networks and attention-based models to enhance classification accuracy by focusing on the most critical features of light curves.  \citet{Dobryakov2021} also successfully applied data-driven methods for Type Ia supernova classification, achieving high performance with minimal manual intervention.

Competitions like the Photometric LSST Astronomical Time-series Classification Challenge (\plasticc, \citealt{ThePLAsTiCCteam2018, Kessler2019a, Hlozek2023}) have significantly advanced the state of the art in transient classification. \plasticc aimed to develop robust classifiers by simulating a non-representative training set for a large, photometrically imbalanced test set. Hosted on Kaggle, this competition saw over 1000 teams employ diverse machine-learning techniques. Building on this, the \rst \romanplasticc simulation \citep{Rose2023} is a comprehensive synthetic dataset designed to simulate the observational conditions expected from the \romanST. It is mainly simulated using \plasticc models \citep{Kessler2019a}, though other more updated models have been used through development. This simulation includes a diverse array of transient astronomical events, each represented by photometric light curves across multiple observational bands. The \romanplasticc simulation's realistic simulation of the Roman Telescope’s observational capabilities ensures that models developed using this data are well-prepared for real-world applications.

One of the most promising approaches to come out of \plasticc was \Parsnip (Parameterized Supernova Identification Pipeline, \citealt{Boone2021}). \Parsnip employs a variational autoencoder (VAE) \citep{Kingma2013} to analyze supernova light curves, generating latent representations that capture the intrinsic features of these astronomical events. There have been previous applications of autoencoders on astronomical light curves \citep{Naul2018, Pasquet2019, Martinez-Palomera2020, Villar2020, Villar2021}, all of which have shown to be effective on tasks such as identifying outliers and classifying photometric light curves. However, as explained in \citet{Boone2021}, these models do not include explicit description of observing symmetries, which leads to the same transient being assigned to more than one location in the latent space if it was being observed under different conditions. \Parsnip's strength comes with adding an explicit \textit{physics} layer on top of the autoencoder, which gives it the ability to create an intrinsic latent space that is invariant to observing conditions like redshift. This makes it particularly effective in handling the diverse and complex nature of transient events. By learning these latent representations, \Parsnip can confidently differentiate between various types of transients, providing accurate classifications even under varying observational conditions.

The primary objective of our study is testing \Parsnip on the simulated \rst \romanplasticc dataset to cluster and classify various transient types based on the model's internal parameters. \cref{sec:dataset} details the dataset characteristics; mainly which types are present and the redshift ranges, along with preprocessing steps we took before using the model. In \cref{sec:parsnip_training}, we provide a brief explanation of the \Parsnip model and its training procedures. \cref{sec:results} presents our results, including performance metrics and classifications done using the \Parsnip classifier; \cref{sec:disc} provides a discussion for the implication of our results and potential areas for future research. Finally, \cref{sec:conc} summarizes the key contributions and outcomes of our study.

\section{Dataset} \label{sec:dataset}

The \rst \romanplasticc dataset is a comprehensive synthetic dataset that is intended to help forecast yields of different classes of transients, train and test classification models, and generally give the community a sense of what the HLTDS can do for their science. It is built on transient models developed by the original \plasticc study detailed in \citet{Kessler2019a}. This dataset accurately simulates the conditions under which the \romanST will operate, allowing researchers to develop and fine-tune models capable of identifying and classifying various transient events. It includes a wide array of transient astronomical events, each represented by a spectral time-series model that enables computing photometric light curves in multiple observational bands. These simulated light curves are degraded with noise and sampled at a five-day cadence to reflect the realistic observational capabilities of the \romanST. They capture the temporal evolution of brightness across different photometric bands, providing a rich dataset for training and validating machine learning models (e.g., \citealt{Qu2021, Aleo2023}).
\subsection{Survey Strategy} \label{sec:survey}
The proposed survey strategy for the Nancy Grace \romanST{}, as outlined by \citet{Rose2021c}, employs a dual-tier observational design to balance broad sky coverage with greater depth in targeted areas. This approach consists of a wide tier, covering approximately 19.04 square degrees using four filters, $F062$, $F087$, $F106$, and $F129$, which span wavelengths from 0.48 to 1.454 microns. This wide tier enables broader sky coverage, capturing a range of transient events across diverse brightness levels. Meanwhile, the deep tier focuses on a narrower area of about 4.2 square degrees, utilizing the filters $F106$, $F129$, $F158$, and $F184$, covering wavelengths from 0.927 to 2.00 microns, as detailed by \citet{Rose2021c}. This setup allows the survey to detect fainter transients with redder coverage. Thus, the survey provides an optimal balance between sky breadth and observational depth. A key feature of this strategy is the five-day cadence maintained throughout the two-year observational period. This cadence offers dense temporal sampling, essential for tracking the evolution of transient events and constructing reliable light curves.

To enhance classification and characterization, the deep tier includes slitless spectroscopy. Observations with a low-resolution prism (R $\sim$ 100), spanning 0.75–1.8 µm, provide additional spectral data for a subset of transients, adding a crucial layer of information that helps differentiate similar types and improves classification accuracy. This spectral data complements the photometric observations, making the dataset particularly valuable for machine learning classification models, as it allows for independent validation and adds a distinct spectroscopic dimension to the dataset.

Field selection criteria include targeting high ecliptic and galactic latitudes to reduce interference from zodiacal light and dust extinction, which enhances data precision and accuracy. The survey strategically overlaps regions with previous or contemporaneous surveys, ensuring continuity and augmenting data richness. This layered observational design not only supports the photometric classification goals of the \romanST but also provides a robust dataset that is highly suited for training machine learning models, offering broad utility across a wide redshift range.

\begin{figure*}[htbp!]
    \centering
    \includegraphics[width=0.35\textwidth]{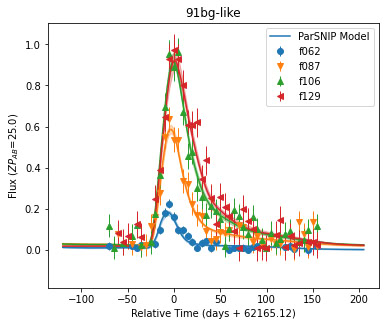}
    \includegraphics[width=0.35\textwidth]{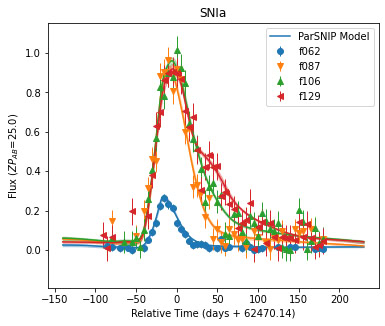}
    \includegraphics[width=0.35\textwidth]{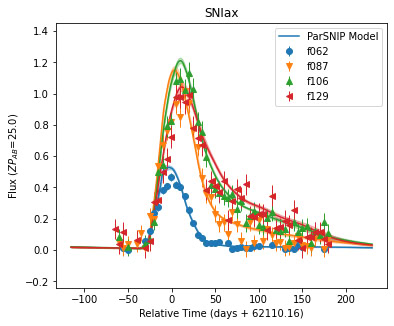}
    \includegraphics[width=0.35\textwidth]{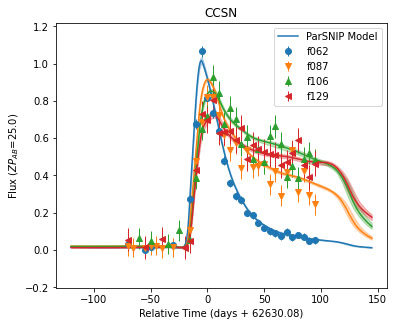}
    \caption{Light curves plotted for each of the four most well-represented transient types: 91bg-like, SNIa, SNIax, and CCSN. Each curve displays flux measurements across multiple Roman broadband filters (\(F062\), \(F087\), \(F106\), \(F129\)) with corresponding \Parsnip model fits superimposed. The diversity in brightness and temporal evolution highlights the differences between transient types and demonstrates the \Parsnip model's ability to capture their unique characteristics.}

    \label{fig:lcs}
\end{figure*}

The observed light curves shown in \cref{fig:lcs} represent the four most well-represented transient types in the dataset: 91bg-like, SNIa, SNIax, and CCSN. These light curves demonstrate the diversity in photometric behavior across different supernova types, as captured by the \Parsnip model. Each panel illustrates the relative flux measurements from multiple Roman broadband filters ($F062$, $F087$, $F106$, $F129$), with the corresponding \Parsnip model fits superimposed. The consistency of the model fits across filters and types highlights the model’s capacity to capture the temporal evolution and spectral energy distribution of transients, even for those with differing luminosity and variability \citep{Leibundgut2000}.

\FloatBarrier
\subsection{Dataset Details} \label{sec:details}

\begin{deluxetable*}{lcccc}[htbp!]
\tablewidth{0pt}
\tablecaption{Counts of transient classes for a representative train and test set. \label{tab:supernova_counts}}
\tablehead{
    \colhead{Type} & \colhead{Train Count} & \colhead{Mean Redshift ($z_\text{train}$)} & \colhead{Test Count} & \colhead{Mean Redshift ($z_\text{test}$)}
}
\startdata
CCSN & 42195 & 1.04 & 4230 & 1.05 \\
SNIa & 15731 & 1.36 & 1539 & 1.36 \\
91bg-like & 1100 & 0.90 & 111 & 0.90 \\
SNIax & 832 & 0.90 & 95 & 0.91 \\
SLSN & 65 & 1.71 & 5 & 2.02 \\
TDE & 36 & 0.75 & 2 & 0.78 \\
PISNe & 9 & 1.54 & 2 & 0.48 \\
ILOT & 32 & 0.43 & 1 & 0.75 \\
\enddata
\end{deluxetable*}
\FloatBarrier
The primary types of transients simulated in the \romanplasticc dataset are Type Ia Supernovae (SNe Ia), Core-Collapse Supernovae (CC~SNe), SNe~Iax, and 91bg-like SNe~Ia, among others, detailed in Table~\ref{tab:supernova_counts}. The dataset initially consisted of around 67,000 transients; however, we had to apply a few cuts in the beginning for transients with less than three phase epochs for transients observed near the beginning or end of the survey. This brought down the number to a little more than 65,000. For each training run, we reserved 90\% for training the \Parsnip model and used the rest for testing and producing classification results (10-fold cross validation). (The test sets were randomly shuffled so only the average fold had exactly 10\%.) Typical counts for each type in both datasets are given in \cref{tab:supernova_counts}.

\begin{figure}[htbp!]
    \centering
    \includegraphics[width=0.8\linewidth, height=2.5in]{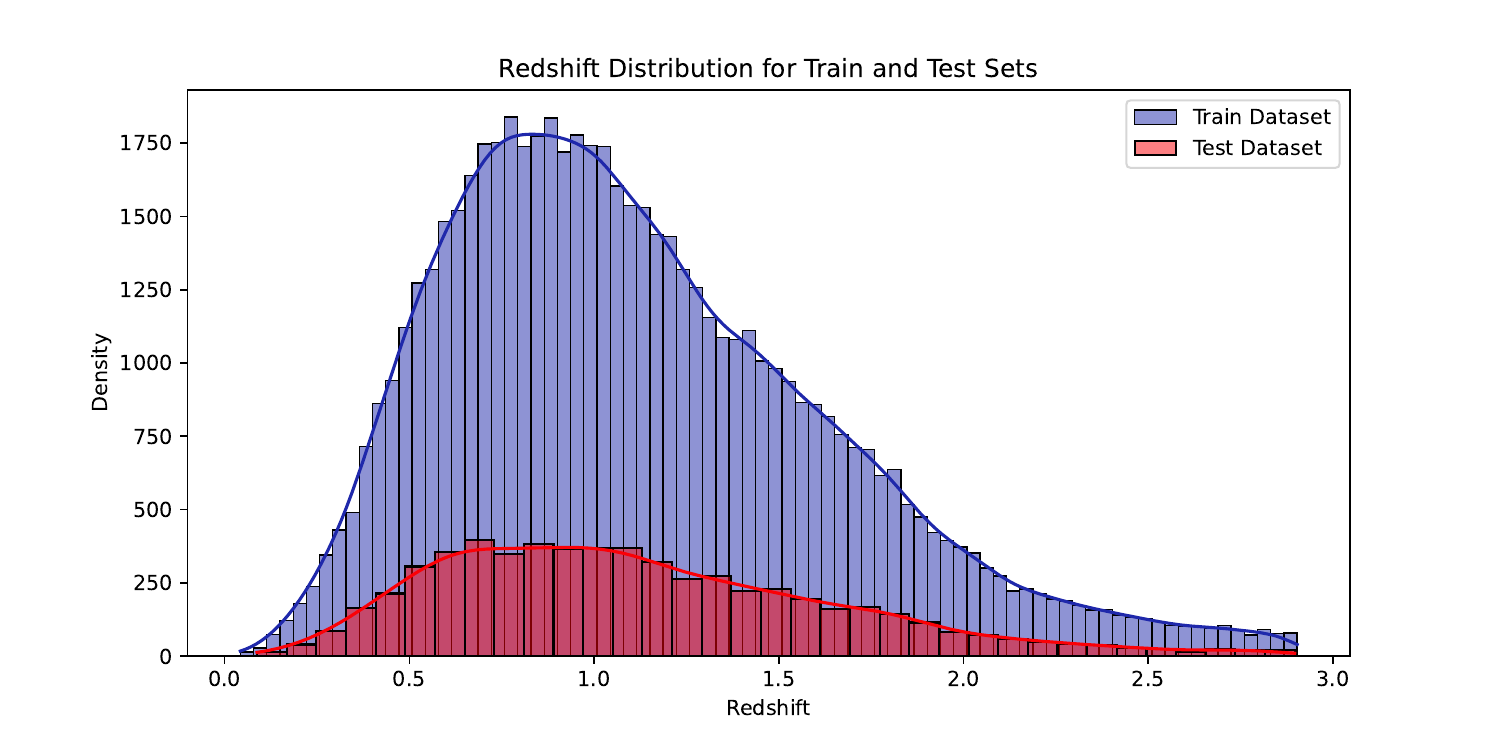}
    \caption{Redshift distribution for representative training and test datasets; note the different binning.}
    \label{fig:z_hist}
\end{figure}

The \rst dataset will span a wide range of redshifts enabling the study of transient events over a significant portion of cosmic history. \cref{fig:z_hist} shows the redshift distribution for a representative training and test dataset. The mean redshift is $\approx$ 1.12 for both the train and test datasets with a noticeable positive skew, giving the best representation for transients with redshifts between 0.5 and 1.5. Since most of the transients in our dataset fall in this range, we expect \Parsnip to have the best performance in this redshift range because it has the best representation of transient classes.  \cref{tab:supernova_counts} also quotes the mean redshift for each class of transient present in our dataset, while \cref{tab:redshift_summary} quotes the same data for the \plasticc dataset used in the original \citet{Boone2021} \Parsnip paper.

\begin{table}[htbp!]
\centering
\begin{tabular}{lrr}
\hline
\textbf{Type} & \textbf{Count} & \textbf{Mean Redshift} \\
\hline
AGN & 101,424 & 0.73 \\
CaRT & 9,680 & 0.26 \\
EB & 96,572 & 0.00 \\
ILOT & 1,702 & 0.16 \\
KN & 133 & 0.14 \\
M-dwarf & 93,494 & 0.00 \\
Mira & 1,453 & 0.00 \\
PISN & 1,172 & 0.88 \\
RRL & 197,155 & 0.00 \\
SLSN-I & 35,782 & 1.51 \\
SNII & 100,150 & 0.41 \\
SNIa & 1,659,831 & 0.56 \\
SNIa-91bg & 40,193 & 0.31 \\
SNIax & 63,664 & 0.39 \\
SNIbc & 175,094 & 0.31 \\
TDE & 13,555 & 0.54 \\
uLens-Binary & 533 & 0.00 \\
uLens-Single & 1,303 & 0.00 \\
\hline
\end{tabular}
\caption{Summary of Type, Count, and Mean Redshift for the PlasTicCC dataset.}
\label{tab:redshift_summary}
\end{table}

\FloatBarrier

\section{Training the \Parsnip Model}
\label{sec:parsnip_training}

We trained the \Parsnip model (lightly updated from \citealt{Boone2021}) on the \rst \romanplasticc dataset to learn latent parameters that represent different transient types. This update consists of time-smoothing regularization as shown in \cref{time_reg}. The model was trained using a variational autoencoder (VAE) neural network architecture, where the input features were photometric light curves, and the output was a set of latent parameters encoding the essential characteristics of each transient.

\subsection{Architecture and Hyperparameters}

As noted above, \Parsnip's architecture incorporates a variational autoencoder (VAE) framework. The key innovation in \Parsnip is its physics layer, which helps ensure that the learned latent space is invariant to observing conditions like redshift. This enables the model to generalize well across different observational environments, improving its ability to differentiate between various transient types. The architecture consists of several components:
\begin{enumerate}
\item Encoder: Maps input light curves to a 3D latent space (called $s_1$, $s_2$, and $s_3$).
\item Decoder: Reconstructs light curves from the latent parameters.
\item Physics Layer: Enforces invariance in the latent space to observational variations, such as redshift and dust extinction.
\end{enumerate}

The model works on a batch size of 128 photometric light curves per batch. The training involved using the \texttt{adam} optimizer which uses a momentum-dependent learning rate that starts at $10^{-4}$ and gets cut by half until it reaches $10^{-6}$, which signals the end of training.

As described in \citet{Boone2021}, multiple regularization techniques are used during the model training. Dropout layers are used inside the neural network to prevent overfitting. Additionally, two regularization terms are added to our loss function, the first is L1 regularization that penalizes the spectra across adjacent bins with a regularization parameter $\eta = 0.001$, this smooths out the spectra and prevents the model from adding high-frequency noise to them, similar to \citet{crenshaw20}. The second regularization term adds an L2 penalty with $\zeta = 7 \times 10^{-5}$ on the photometric light curves across adjacent phases, this also provides smoother light curves while minimizing noise in them. The value for $\zeta$ was decided upon after the hyperparameter tuning phase in which we tried an array of different numbers ranging from $10^{-5}$ to $5 \times 10^{-4}$. This last regularization parameter was manually added by us during this study and was not present in the original work of \citet{Boone2021}. 

\begin{equation} \label{flux_reg}
   (\eta = 0.001) \sum_i \bigg( \frac{ d_{\text{int},\boldsymbol{\theta},n}(\boldsymbol{t},\boldsymbol{s}_i) - d_{\text{int},\boldsymbol{\theta},n+1}(\boldsymbol{t},\boldsymbol{s}_i)}{d_{\text{int},\boldsymbol{\theta},n}(\boldsymbol{t},\boldsymbol{s}_i) + d_{\text{int},\boldsymbol{\theta},n+1}(\boldsymbol{t},\boldsymbol{s}_i)} \bigg)^2
\end{equation}

\begin{equation} \label{time_reg}
        (\zeta = 7 \times 10^{-5}) \sum_i \left( d_{\text{int},\boldsymbol{\theta},n}(\boldsymbol{t}+1,\boldsymbol{s}_i) + d_{\text{int},\boldsymbol{\theta},n}(\boldsymbol{t}-1,\boldsymbol{s}_i) -
    2d_{\text{int},\boldsymbol{\theta},n}(\boldsymbol{t},\boldsymbol{s}_i)\right)^2 \\ 
\end{equation}

\subsection{Comparison Between Roman and Rubin Telescopes}
The ParSNIP model was initially developed to improve transient classification for the Rubin Observatory’s Legacy Survey of Space and Time (LSST) dataset by addressing the challenge of separating intrinsic diversity in transient light curves from observational biases, such as redshift and dust effects. With its innovative physics-based layer, ParSNIP manages these observational “symmetries” effectively, creating redshift-invariant representations that support reliable classifications across diverse observing conditions. For the Rubin dataset, ParSNIP showed remarkable improvements over traditional classification methods by maintaining the physical characteristics of transients even under varying observational conditions like redshift and signal noise, enabling it to perform well within LSST’s optical bands ($u$, $g$, $r$, $i$, $z$, $y$) \citep{Boone2021}.

In adapting ParSNIP for the Roman Space Telescope, it allows us to test the effects of differences in survey parameters such as cadence, redshift distribution, and wavelength range. While the LSST operates in the optical range with a twice-weekly cadence across half the sky, Roman’s high-latitude survey takes a different approach: it uses a five-day cadence over two years to achieve greater temporal sampling and depth across a smaller field. This enhanced cadence allows Roman to detect fainter transients that may only be observable in the NIR bands (spanning $F062$ to $F184$), demanding adjustments to ParSNIP’s intrinsic representation model \citep{Boone2021}. These adaptations ensure that ParSNIP can preserve classification accuracy within Roman’s deeper, more temporally-rich dataset, which spans a wavelength range of $0.48$ to $2.00$ microns and captures a broader array of transient events.

\section{Results and Classifications}
\label{sec:results}

\subsection{Latent Spectra Analysis}
\begin{figure*}[htbp!]
    \centering
    \includegraphics[width=0.35\textwidth]{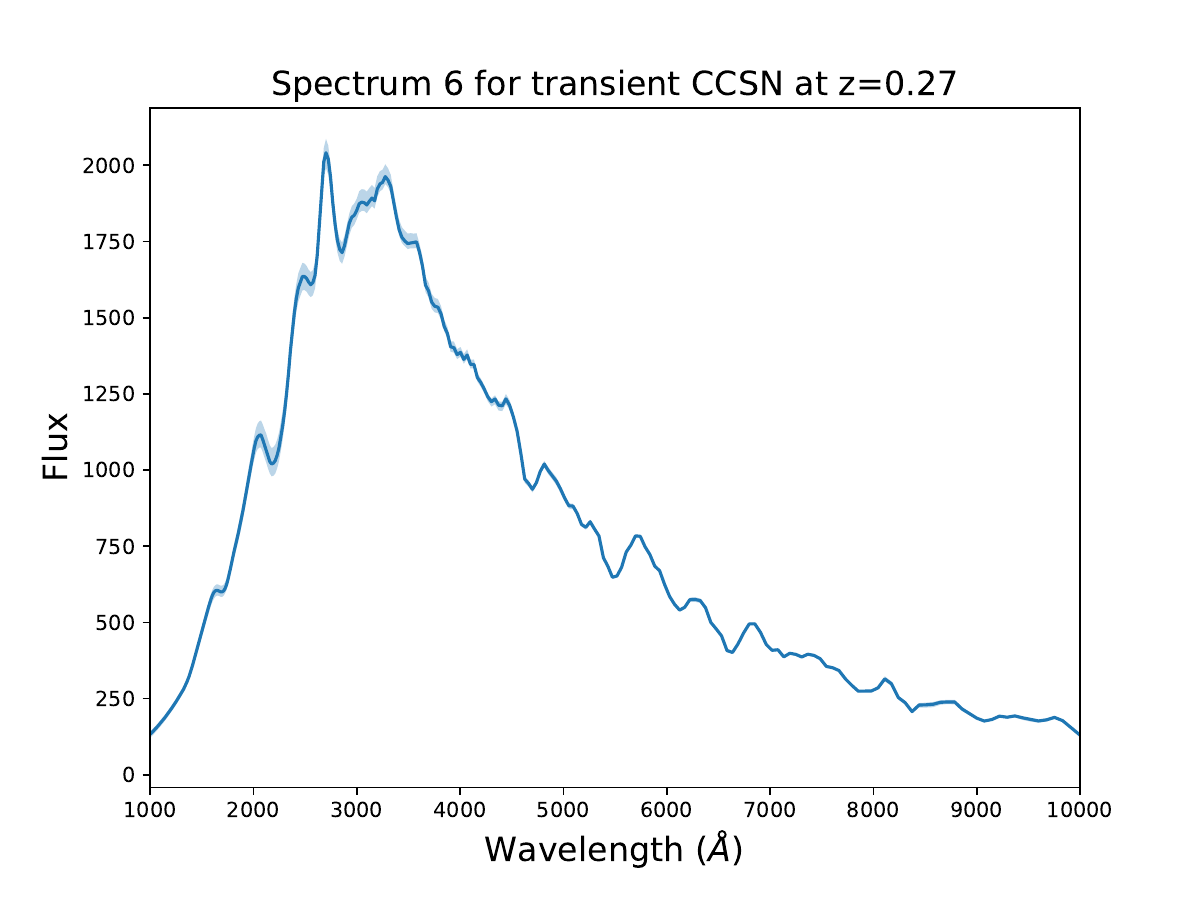}
    \includegraphics[width=0.35\textwidth]{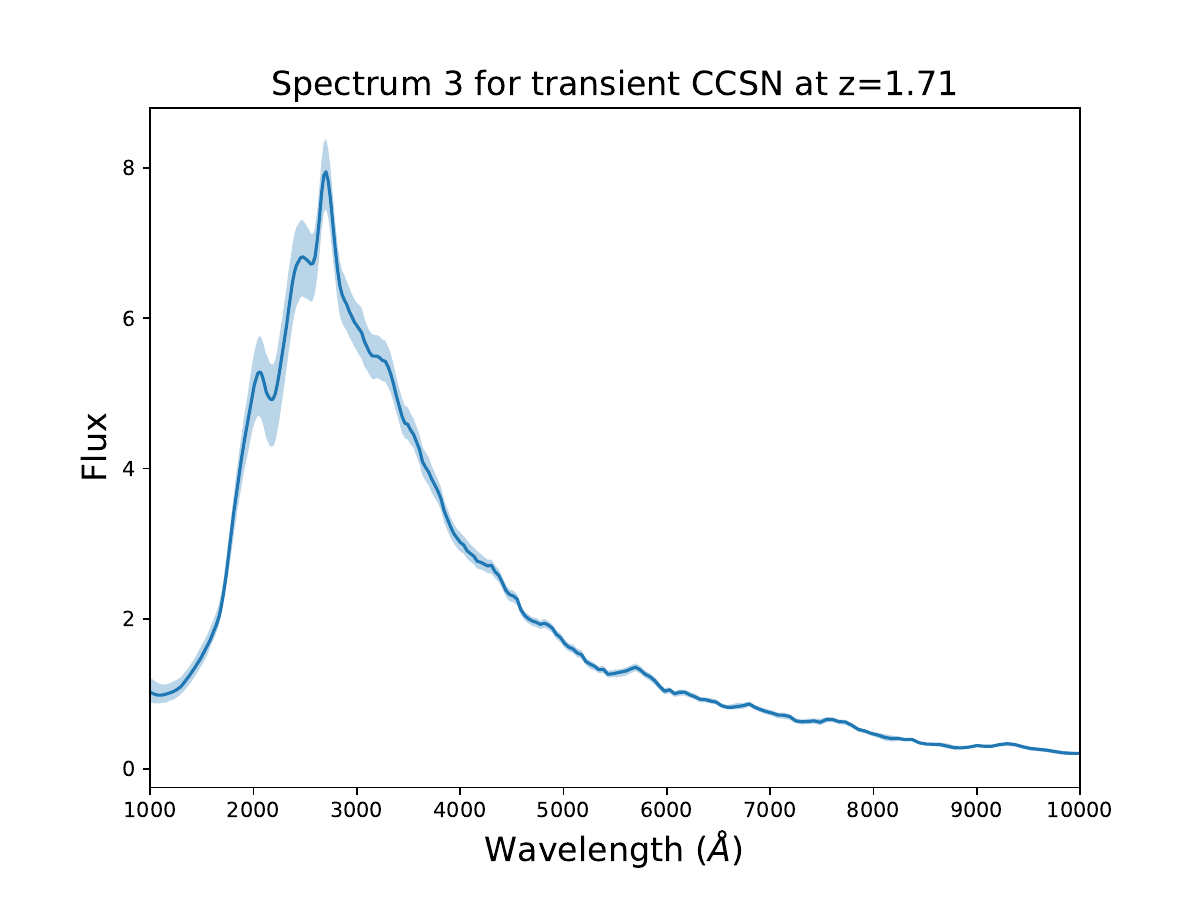}
    \includegraphics[width=0.35\textwidth]{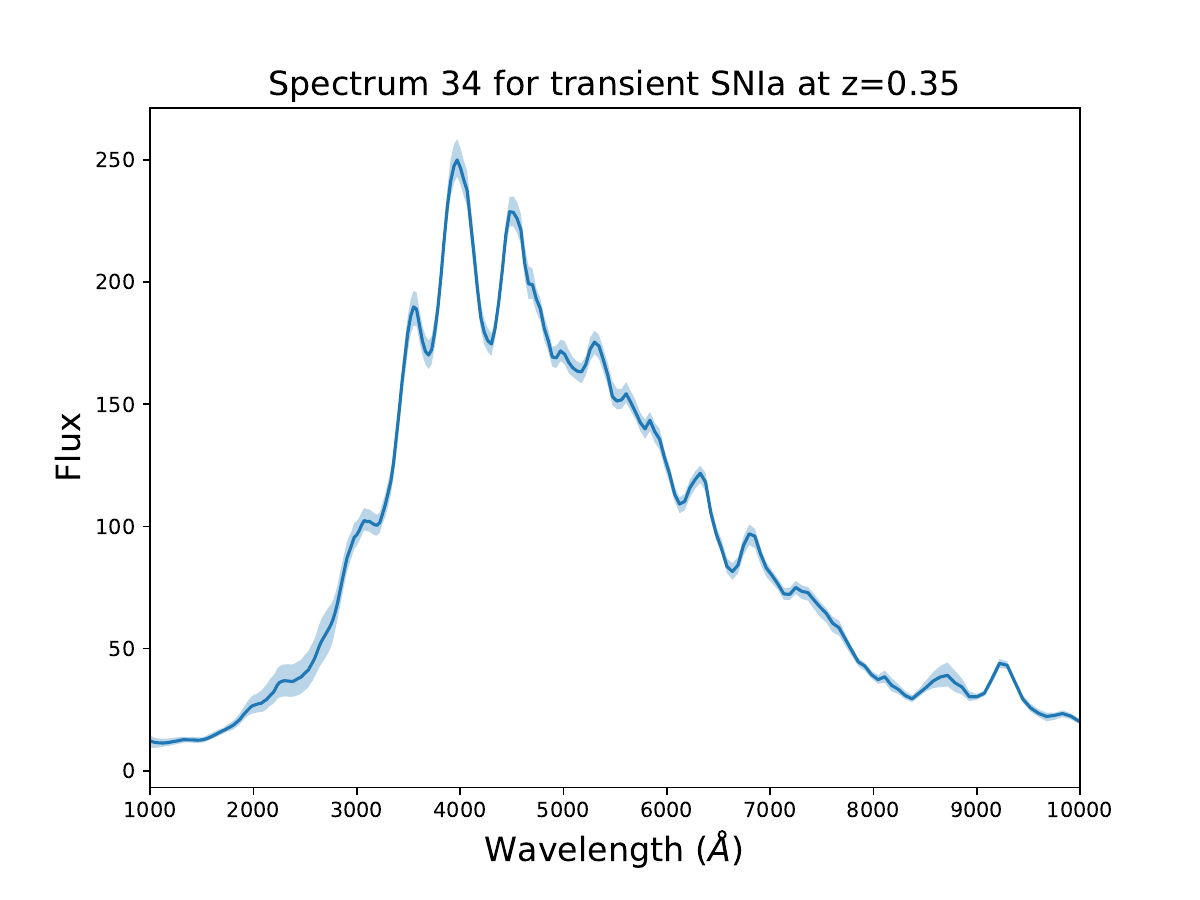}
    \includegraphics[width=0.35\textwidth]{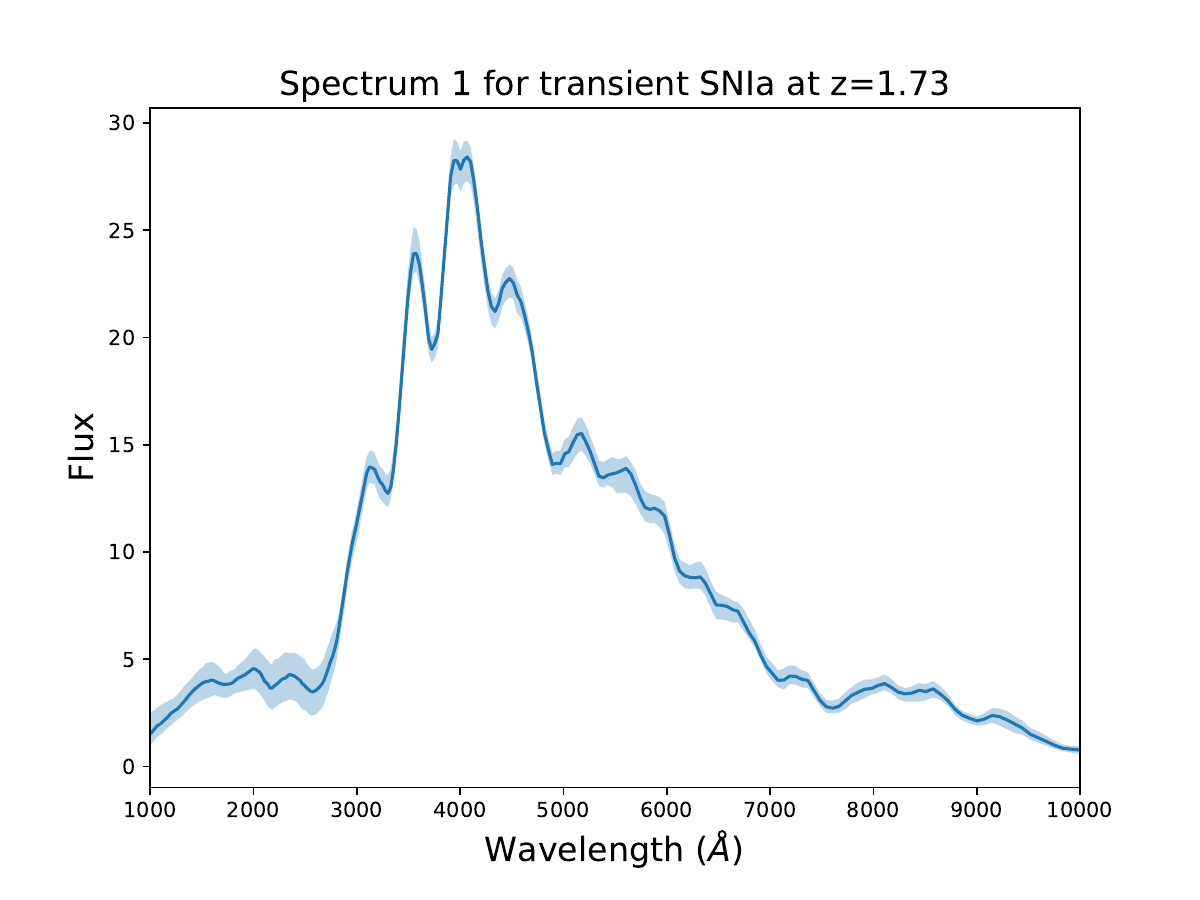}
    \includegraphics[width=0.35\textwidth]{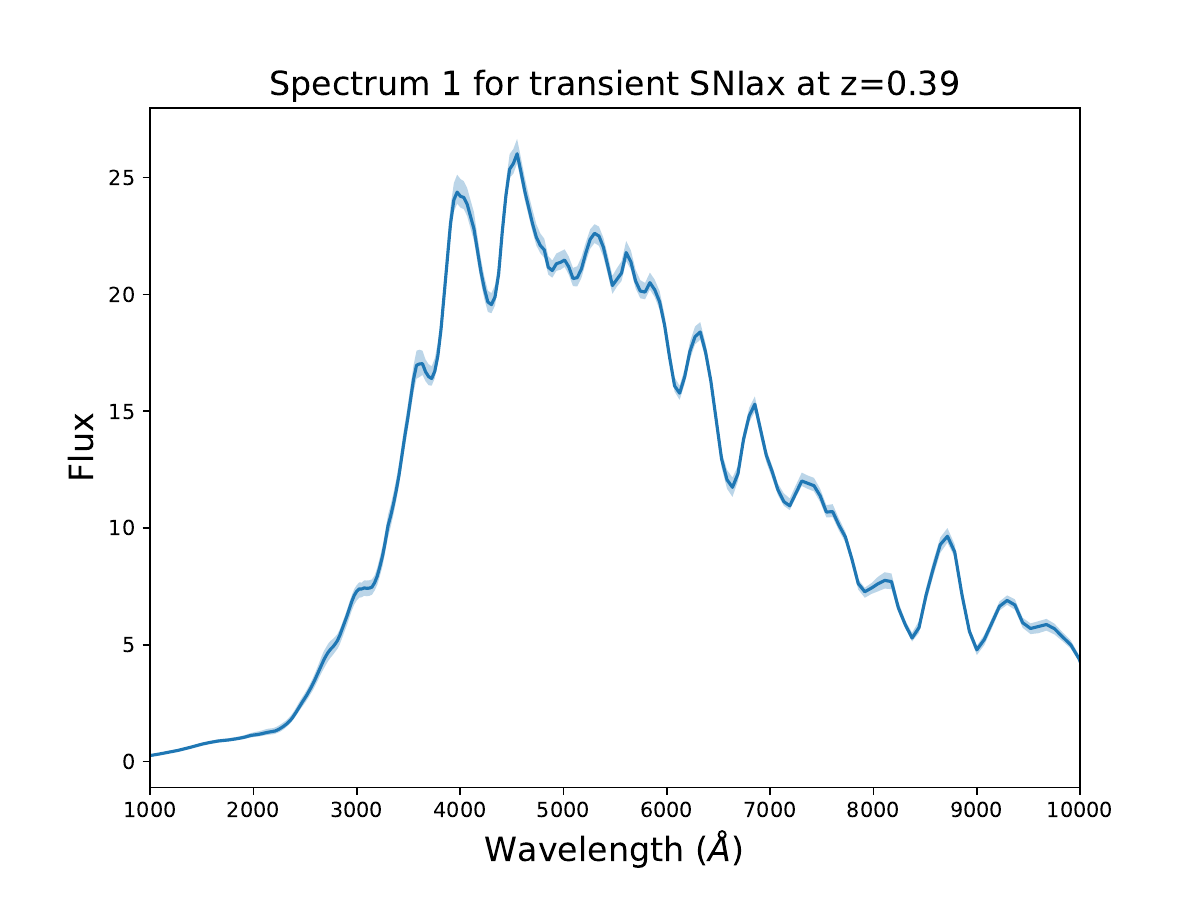}
    \includegraphics[width=0.35\textwidth]{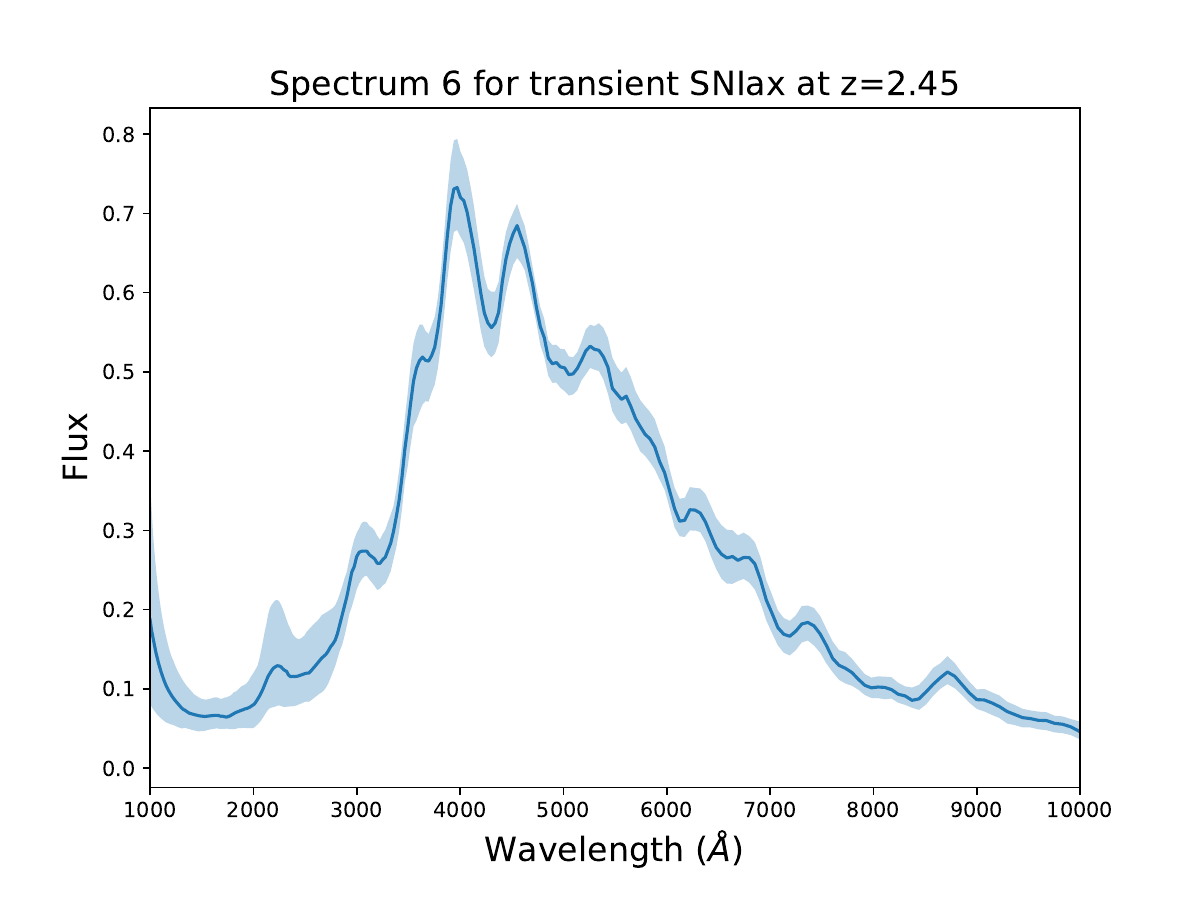}
    \includegraphics[width=0.35\textwidth]{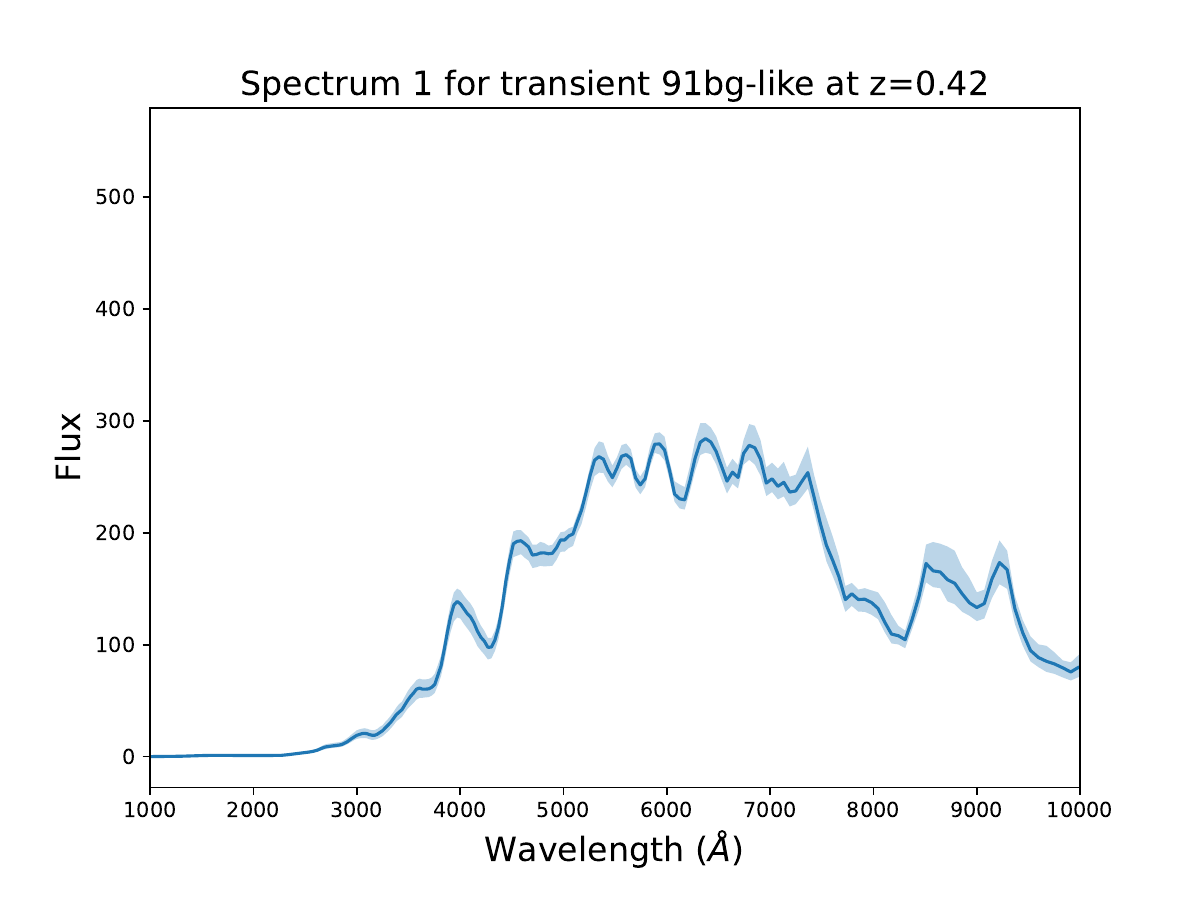}
    \includegraphics[width=0.35\textwidth]{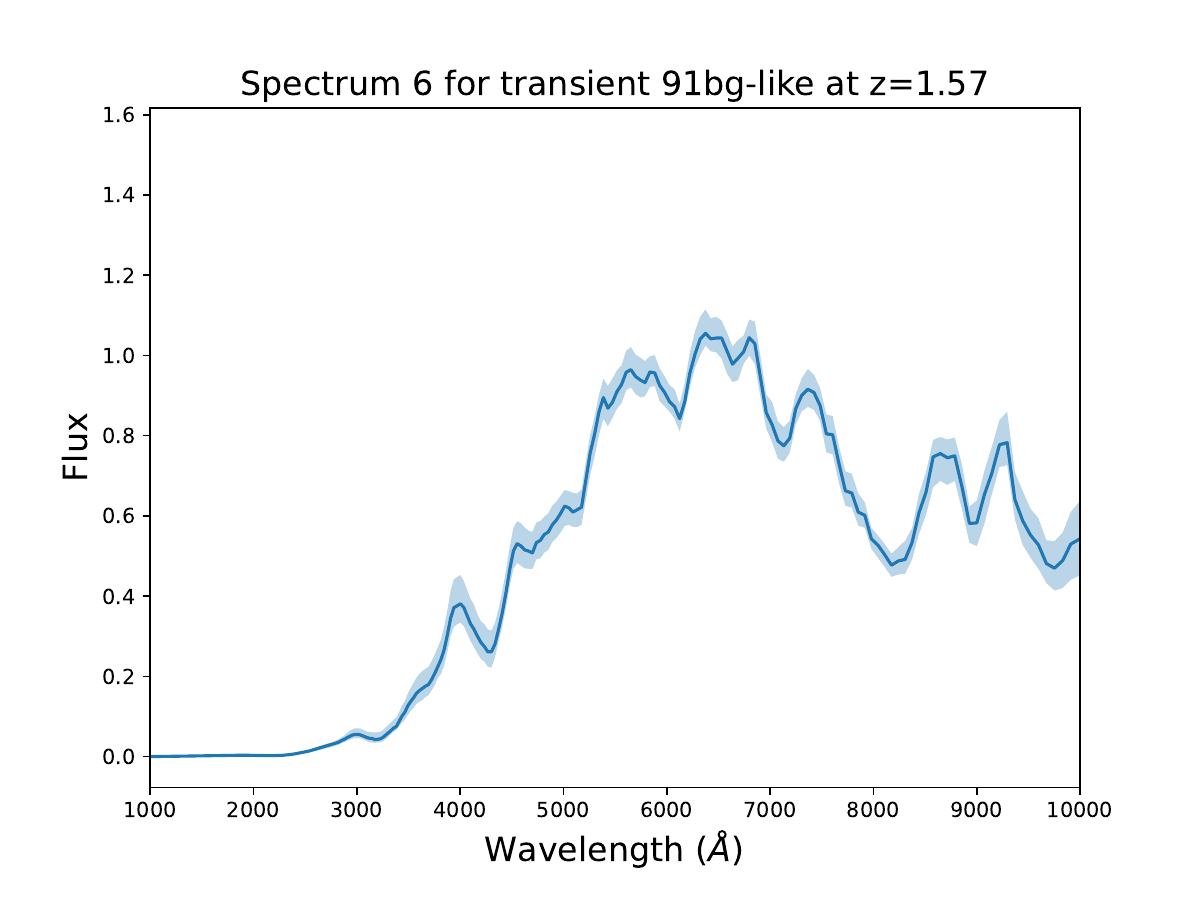}
    \caption{Latent spectra of different transient types across low and high redshift ranges. The left panel shows spectra for Core-Collapse Supernovae (CCSN), Type Ia Supernovae (SNIa), peculiar Type Iax Supernovae (SNIax), and 91bg-like supernovae at low redshifts (\( z \sim 0.27 \) to \( z \sim 0.42 \)). The right panel displays the corresponding spectra at high redshifts (\( z \sim 1.03 \) to \( z \sim 2.45 \)). The spectra illustrate the ability of the \Parsnip model to preserve key spectral features across redshifts, demonstrating its redshift-invariant representation of transient types.}
    \label{fig:spectra}
\end{figure*}

After obtaining the latent \Parsnip parameters for each event, we plotted the latent spectra decoded for every type. The latent spectra generated by the \Parsnip model demonstrate the ability to preserve critical spectral characteristics of various transient types across a wide range of redshifts. This capability is essential for identifying and analyzing supernovae in cosmological datasets, where redshift-induced changes in observed wavelength can obscure key features. As illustrated in \cref{fig:spectra}, each transient type retains its defining spectral features in the latent space, underscoring the model’s effectiveness in encoding intrinsic properties without introducing distortions across redshifts.

Core-Collapse Supernovae (CCSN) provide a clear example of the model’s ability to preserve critical spectral features across redshifts. As shown in \cref{fig:spectra}, hydrogen Balmer lines and oxygen profiles, which are characteristic of CCSN spectra, remain visible at both low (z=0.27) and medium redshift (z=1.03). Despite the expected shifts in wavelength due to redshift, these features are distinct across the latent spectra, highlighting the model’s effectiveness in retaining key CCSN markers even as distance increases \citep{Filippenko1997}. The presence of these spectral indicators in the latent space supports the applicability of the \Parsnip model for studies focused on accurately identifying CCSN in cosmological datasets with significant redshift coverage.

Type Ia (SNIa) and peculiar Type Iax (SNIax) supernovae further demonstrate the model’s capability to maintain essential spectral features across redshifts. As shown in \cref{fig:spectra}, SNIa spectra consistently retain their characteristic Si II absorption near 6150 Å and calcium lines in the blue region, which are critical for their role as standard candles in cosmology \citep{Nugent1995}. Similarly, SNIax spectra preserve broader silicon and carbon features, as well as increased spectral diversity, at both low (z=0.39) and high redshift (z=2.45). These latent spectra effectively capture the defining spectral properties of SNIa and SNIax, ensuring their reliable identification across a wide redshift range \citep{Foley2013}.

The spectra of 91bg-like emphasizes the ability of the latent space to represent the diversity of transient types. For example, broad iron-group absorption and weaker silicon features—characteristic of this dimmer subtype—are clearly visible at low (z=0.42) and high redshift (z=1.57), as shown in \cref{fig:spectra} \citep{Dhawan2011}. These spectra highlight the model’s precision in encoding high-luminosity transients and preserving their defining features across cosmological distances.

Overall, the \Parsnip model successfully encodes the intrinsic spectral properties of transient types in its latent space, ensuring that their defining features are preserved across redshifts. As illustrated in \cref{fig:spectra}, this capacity allows for accurate classification and analysis of supernovae in datasets with significant redshift coverage. The preservation of these key spectral markers not only facilitates transient identification but also supports broader astrophysical investigations into supernova populations and their evolution.

Furthermore, we visualize the latent space to ensure good separation of transient classes across redshift, train a LightGBM decision tree on those parameters for photometric classification, and evaluate the classification performance.

\subsection{Latent Parameter Visualization}

\begin{figure}[htbp!]
    \centering
    \includegraphics[width=\textwidth, height = 2in]{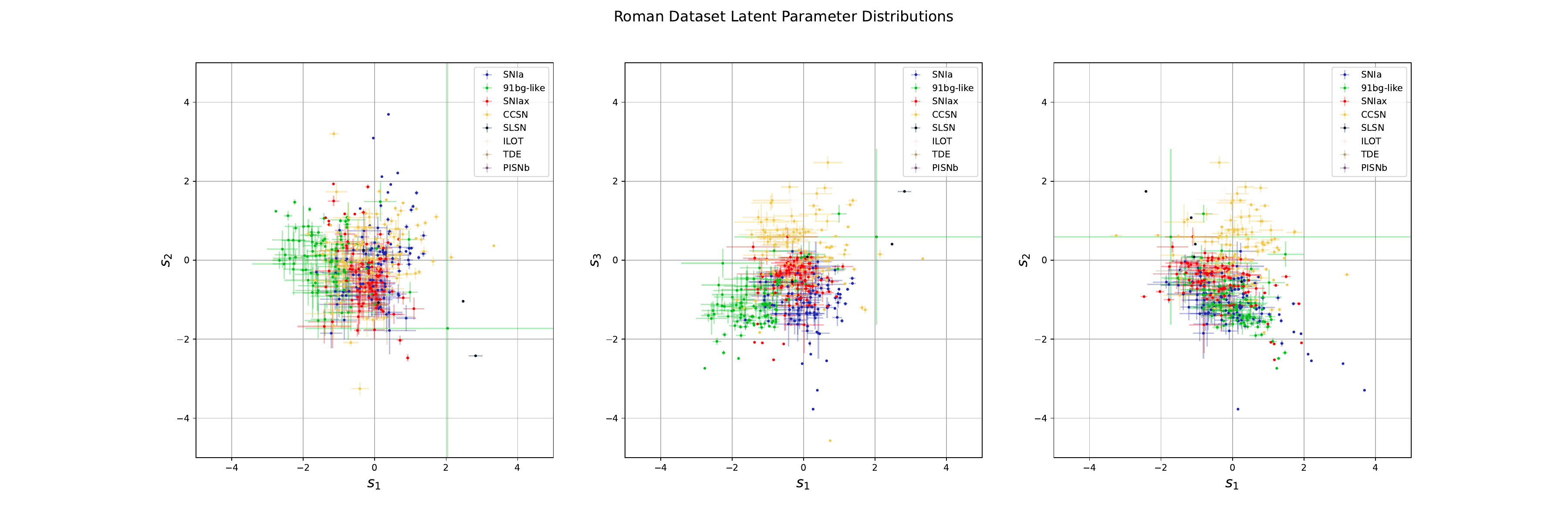}
    \caption{Clustering the data by latent parameters showing the trained model's ability to differentiate between many transient types.}
    \label{fig:RomanClassification}
\end{figure}

To evaluate the \Parsnip model’s effectiveness in distinguishing transient types, we visualized the latent parameters ($s_1 , s_2, s_3$) it learned during training. These parameters capture intrinsic features of the light curves, and projecting them into a two-dimensional space allows us to assess how well the model grouped similar transient types together. In Figure~\ref{fig:RomanClassification}, we see that the latent parameters form distinct clusters for different transient types, including SNe~Ia, SNe~Iax, and CC~SNe. This ability to maintain separation between types is critical for accurate classification, especially in distinguishing between closely related types, such as normal SNe Ia and 91bg-like SNe Ia. The reasonably clear separation between clusters shows that \Parsnip can effectively represent the diversity of transient events in the dataset.

\begin{figure*}[htbp!]
    \centering
    \includegraphics[width=0.9\textwidth]{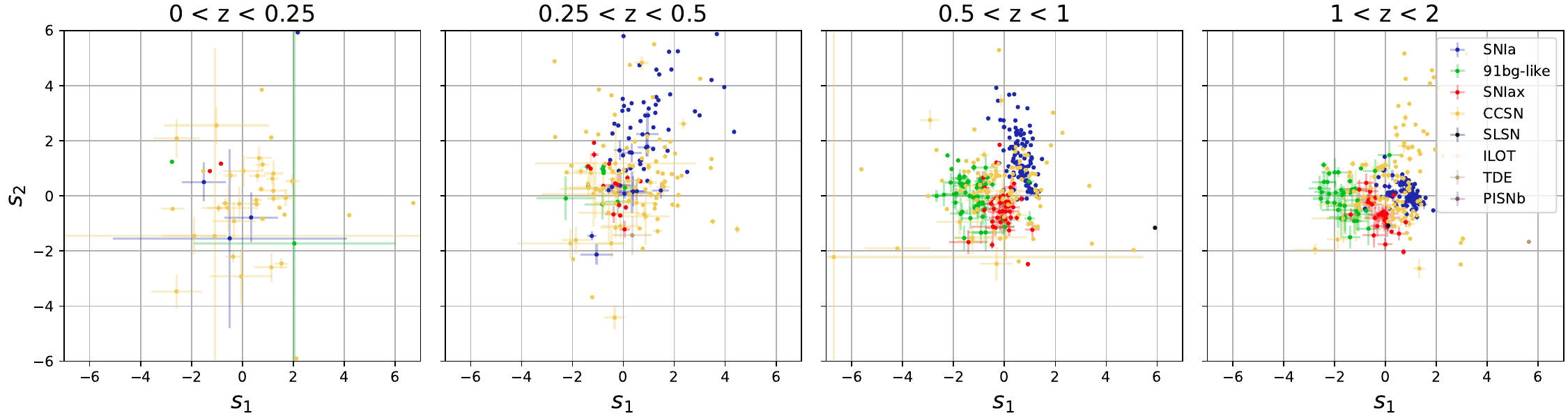}
    \includegraphics[width=0.9\textwidth]{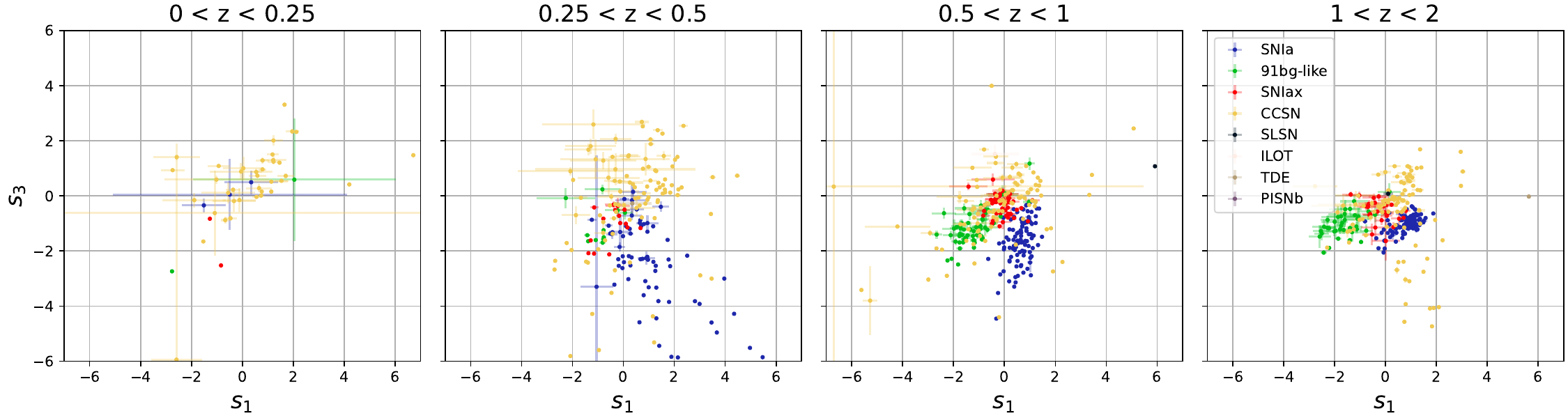}
    \includegraphics[width=0.9\textwidth]{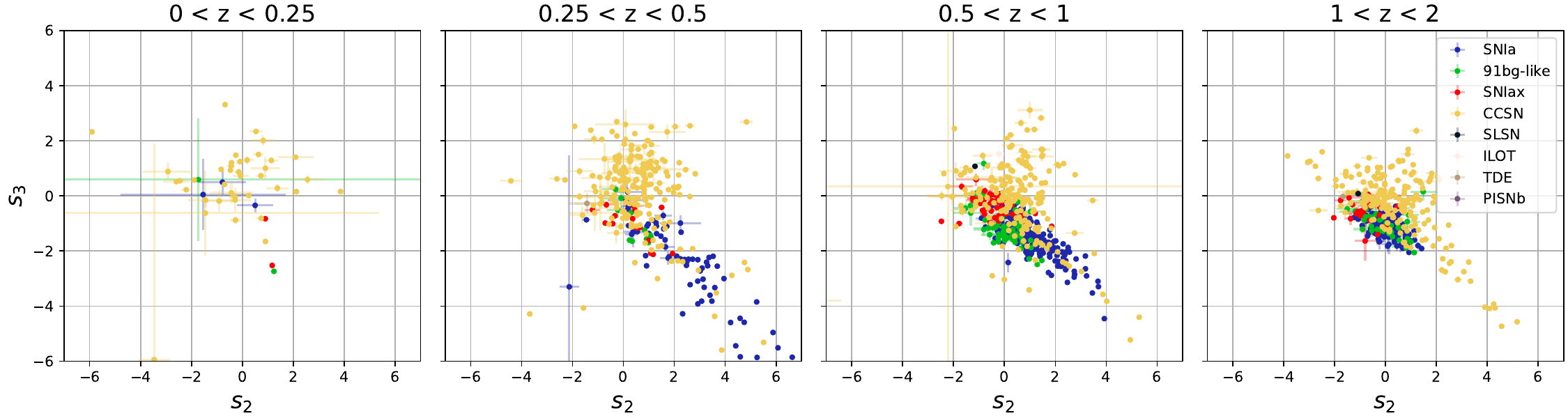}
    \caption{Scatter plots of latent parameters  $s_1 , s_2, s_3$ for different transient types in various redshift bins. Each panel represents a different redshift range, with the x-axis and y-axis showing the two latent dimensions. Different colors and markers represent different transient types. The distinct clustering patterns indicate that the latent parameters effectively capture the characteristics of each transient type. }
    \label{fig:LatentClusters}
\end{figure*}

Figure~\ref{fig:LatentClusters} shows a similar plot in redshift bins. The clustering is consistent across different redshifts, demonstrating that the latent parameters are robust and invariant to observational conditions like redshift.

\FloatBarrier %
\subsection{Latent Space Analysis}
The \textit{Davies-Bouldin Index} (DBI) is a commonly used metric for evaluating clustering quality, where lower values indicate more compact and well-separated clusters. The DBI assesses both intra-cluster similarity and inter-cluster separation, with a lower DBI suggesting clusters are internally cohesive and distinct from one another. This metric is particularly useful for identifying compact clusters without strictly penalizing clusters that fall close together, making it well-suited for latent space analysis in our dataset \citep{Davies1979}.

Our DBI analysis reveals notable differences in clustering quality across redshift bins and parameter combinations. Overall, the combination of \( s_1 \) and \( s_3 \) yields the lowest holistic DBI (2.39), suggesting it provides the most compact and well-defined clustering structure across the entire dataset. Similarly, the three-parameter combination \( s_1, s_2, s_3 \) achieves a comparable holistic DBI (2.48), indicating that including all three parameters supports compact clustering without significantly increasing complexity. In contrast, the combination of \( s_1 \) and \( s_2 \) has a higher holistic DBI (6.74), suggesting weaker clustering quality, particularly influenced by a high DBI in the 0–0.5 redshift bin (34.64), where clusters are less distinct.

Examining the redshift bins individually, the 0.5–1 and 1–1.5 bins display the lowest DBI values across all parameter combinations, indicating that clustering is more distinct in these mid-redshift ranges. For instance, the \( s_1, s_3 \) combination achieves a DBI of 1.39 in the 0.5–1 bin and 1.45 in the 1–1.5 bin, which are among the lowest across all bins and combinations. Conversely, clustering quality decreases at both the lower (0–0.5) and higher (1.5–2 and 2–\(\infty\)) redshift ranges, where DBI values are relatively higher, likely due to increased overlap or less distinct separation in these redshift regimes. This analysis suggests that the latent parameter space captures transient characteristics effectively in the intermediate redshift bins but that clustering quality diminishes at lower and higher redshift extremes, possibly due to increased observational noise or intrinsic variability in the transients.

The \textit{Calinski-Harabasz (CH) Index}, introduced by \citep{Calinski1974}, aligns closely with the \textit{Davies-Bouldin Index (DBI)} findings, further supporting the clustering analysis across different parameter combinations and redshift bins. Higher CH Index values indicate more compact and well-separated clusters, particularly for the combinations \( s_1, s_3 \) and \( s_2, s_3 \), which consistently show strong clustering performance in mid-redshift ranges. For example, the combination \( s_1, s_3 \) has the highest holistic CH Index (319.37) and displays notably high values in the 0.5–1 (183.84) and 1–2 (194.62) redshift bins. This supports the DBI results, which also indicated that this parameter combination provides compact clustering at mid-redshifts. Similarly, the three-parameter combination \( s_1, s_2, s_3 \) achieves a holistic CH Index of 190.97 and relatively high values in the 0.5–1 (116.09) and 1–2 (127.26) bins, reinforcing the DBI findings that clustering quality is well-maintained when all three parameters are included, particularly in these intermediate redshift ranges.

The clustering performance indicated by both CH and DBI suggests that certain parameter combinations, such as \( s_1, s_3 \), are especially effective at distinguishing latent space representations across mid-redshift bins. This clustering behavior is critical for accurately categorizing and analyzing transient types, as it impacts the subsequent training of the \Parsnip classifier. By focusing on parameter combinations that maximize compact clustering, we enhance the classifier's ability to identify distinct transient classes in mid-redshift ranges. The following section outlines the steps involved in training the \Parsnip classifier, incorporating the clustering insights gained to optimize classification accuracy across the dataset.

\subsection{Training the \Parsnip Classifier}

Following the cluster analysis, a LightGBM decision tree \citep{ke17} classifier was also trained on the \rst \romanplasticc dataset to predict the transient types based on the learned latent parameters. The classifier's performance was evaluated by comparing its predictions with the actual types in the dataset. We show two types of results: 1) attempting to classify each transient type, 2) a one-vs-all classification attempting to classify each event as belonging to a specific class vs any other class. It is this latter type that is most important for SN~Ia cosmology, i.e., distinguishing normal SNe~Ia from all other classes. The classifier uses 10-fold cross-validation to ensure robust performance. The features used for classification are color, luminosity, the three latent parameters ($s_1$, $s_2$, $s_3$), and the uncertainty measurements associated with these parameters, along with the uncertainty in the reference time.

The classifier training involves the following steps:\\ 1. \textbf{Feature Extraction}: The relevant features are extracted from the predictions generated by the \Parsnip model. These features include color, luminosity, and their respective errors. \\
2. \textbf{Label Assignment}: The true labels for each light curve are assigned. If a specific target label is provided, a one-vs-all classification is performed; otherwise, a multi-class classification is conducted. \\
3. \textbf{Cross-Validation}: The dataset is divided into 10 folds using stratified K-fold cross-validation, ensuring that augmentations of the same object remain in the same fold. This helps in maintaining the consistency and reliability of the model. \\
4. \textbf{Model Training}: For each fold, a LightGBM model is trained using the training subset, with weights adjusted to normalize class counts. The classifier’s hyperparameters are set to optimize binary or multi-class classification, depending on the context. The classifier attempts to optimize a weighted multi-log loss metric, which assigns weights to each class to address class imbalances, similar to the approach used in (\plasticc) \citep{Kessler2019a}. This approach makes it suitable for imbalanced datasets like ours. \\
5. \textbf{Out-of-Sample Predictions}: For each fold, the classifier is trained on nine folds and makes out-of-sample predictions on the remaining fold. The final classification probabilities are then obtained by averaging the predicted probabilities across all folds for each light curve.
\\

\subsection{Classification}

\begin{figure}[htbp!]
    \centering
    \includegraphics[scale=0.55]{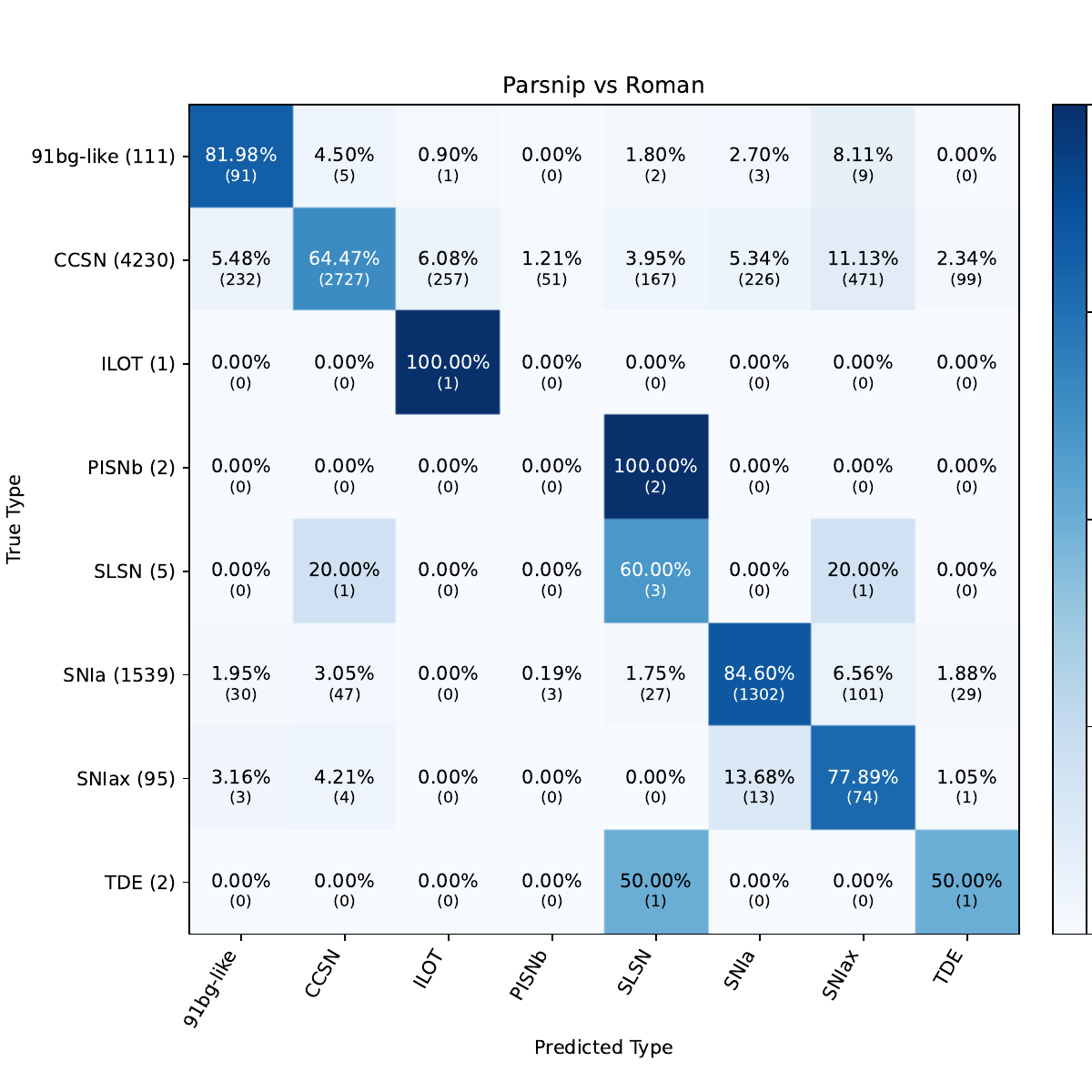}
    \caption{Confusion matrix showing the performance of the \Parsnip classifier on the \rst dataset. The rows represent the true transient types, and the columns represent the predicted types. The diagonal elements indicate the number of correctly classified instances for each type, while the off-diagonal elements represent misclassifications. The high values along the diagonal demonstrate the classifier’s accuracy in predicting transient types. 
    \label{fig:ConfusionMatrix}}
\end{figure}

The classifier’s performance was assessed using confusion matrices and receiver operating characteristic (ROC) curves. As shown in Figure~\ref{fig:ConfusionMatrix}, the confusion matrix shows that the classifier performs well overall, particularly for well-represented types such as SNe Ia, SNe Iax, and CCSNe. Misclassifications occurred mostly with less common types like SLSNe and PISNe, reflecting the imbalanced nature of the dataset. It is also worth noting that all PISNe has been misclassified as SLSN, this is due to their similarity and the fact that the dataset had significantly more representation for the latter class of supernovae.

\FloatBarrier %

\begin{figure*}[htbp!]
    \centering
    \includegraphics[width=\textwidth]{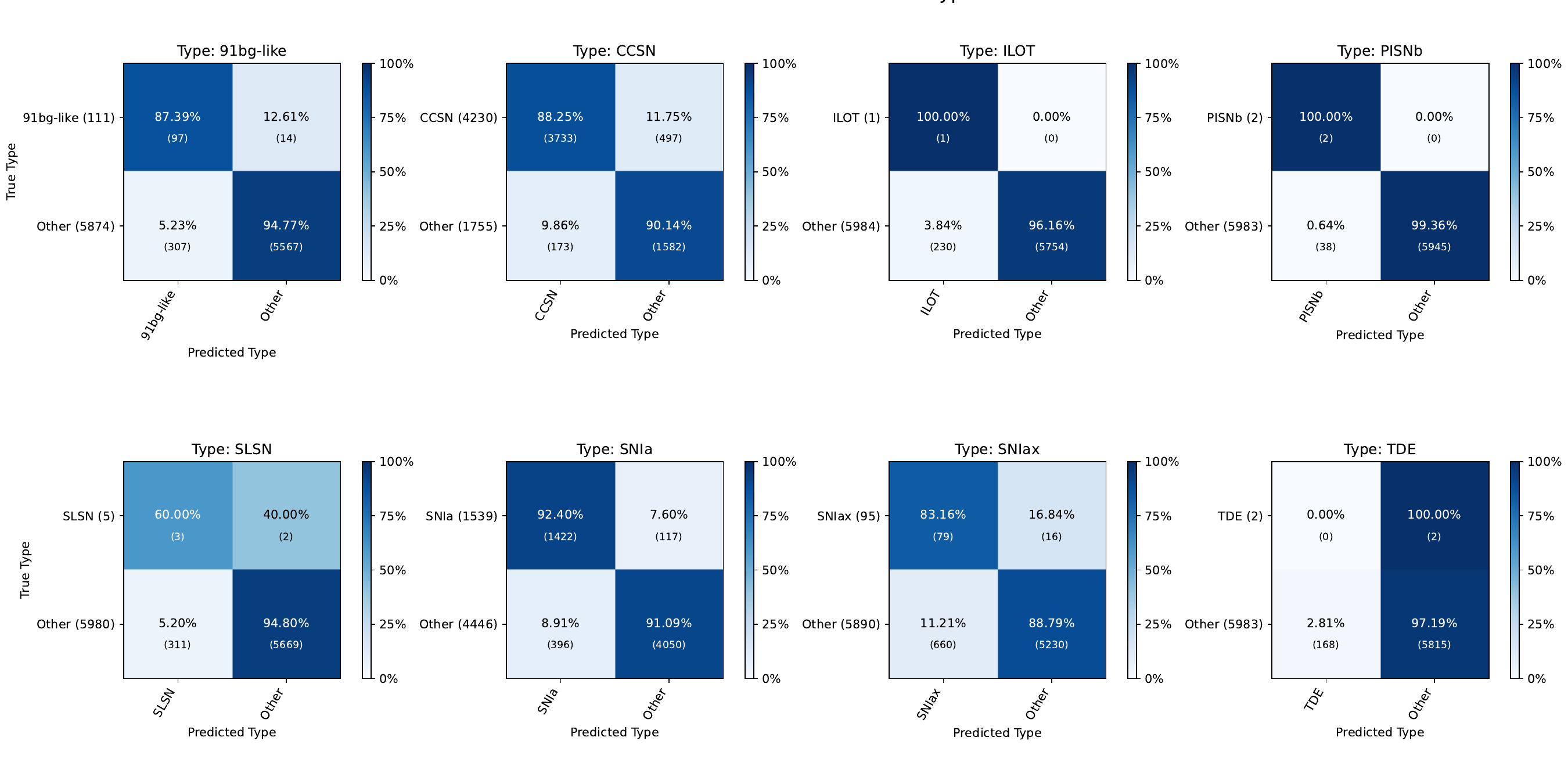}
    \caption{Confusion matrices for the \Parsnip classifier across various transient types. Rows represent true types, and columns represent predicted types. The strong diagonal elements reflect high classification accuracy, with off-diagonal elements indicating areas for potential improvement. SNIa, 91bg-like and CCSN get great results of over 85\% for all of them, while ILOT,TDE,PISNe get mixed results due to insufficient represtnation.}
    \label{fig:ConfusionMatrixSC}
\end{figure*}

In addition to the multi-class classification, we trained single-class classifiers that distinguished each transient type (e.g., SNe~Ia) from all others, relevant for analyses that have to pick out one type specifically (as in SN~Ia cosmology). This approach also produced promising results, with precision and accuracy increasing as the representation of a given type grew. The confusion matrices provide a detailed overview of the classifier’s performance, revealing areas where certain transient types are often confused with one another. This granular view, shown in Figure~\ref{fig:ConfusionMatrixSC}, is key for understanding which transient types present classification challenges and where the model can be improved. Generally, the classifier works quite effectively when the transient type has sufficient representation in the dataset, such as SNe~Ia, CC~SNe, and SNe~91bg-like. SNe with low representation exhibit mixed outcomes, such as the 100\% for ILOT and TDE, or the 50\% shown for PISNe, none of these types were represented by more than three entries in a typical test dataset.

\FloatBarrier

\begin{figure*}[htbp!]
    \centering
    \includegraphics[width = \textwidth]{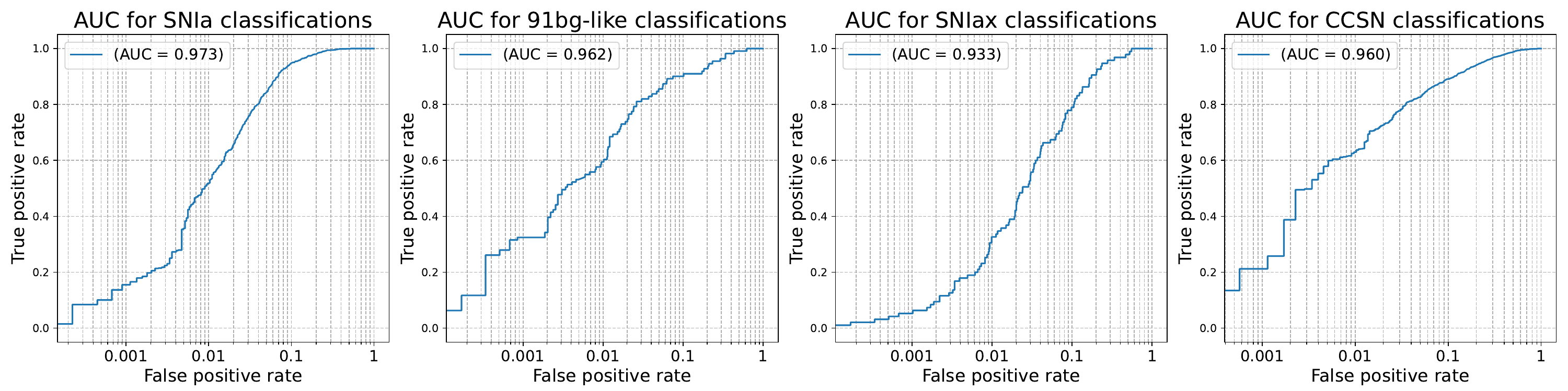}
    \caption{ROC curves for different transient types using the \rst classifier. Each panel shows the ROC curve for a specific transient type, with FPR on a log scale (x-axis) and TPR on a linear scale (y-axis). AUC values in the legends highlight classifier performance, with strong results for well-represented types like SNIa, achieving TPR higher than 95\% at 10\% FPR, and similarly high performance for CCSN and 91bg-like types.}
    \label{fig:ROCCurveSC}
\end{figure*}

To better evaluate our classifier, ROC curves for all transient types are shown in \cref{fig:ROCCurveSC}, providing a detailed assessment of performance across redshift bins. The ROC curves plot True Positive Rate (TPR) against False Positive Rate (FPR) for one-vs-all classification, differing from the multi-class confusion matrices in \cref{fig:ConfusionMatrixSC} and \cref{fig:ConfusionMatrix}. FPR is shown on a logarithmic scale to evaluate the classifier's ability to distinguish true positives from false positives across thresholds. Each subplot represents the ROC curve for a specific transient type, highlighting the classifier's discriminative power.

The AUC (Area Under the Curve) values summarize the classifier's performance, with higher AUC values indicating better discrimination. For example, the ROC curve for SNIa shows a high AUC, reflecting strong classification performance, comparable to the AUC of 0.977 for SNe~Ia in the larger \plasticc dataset \citep{Boone2021}. Lower AUC values for other types, such as CC~SNe, highlight areas for improvement in classification accuracy.

\FloatBarrier

\begin{figure*}[htbp!]
    \centering
    \includegraphics[width=\textwidth, height=3.2in]{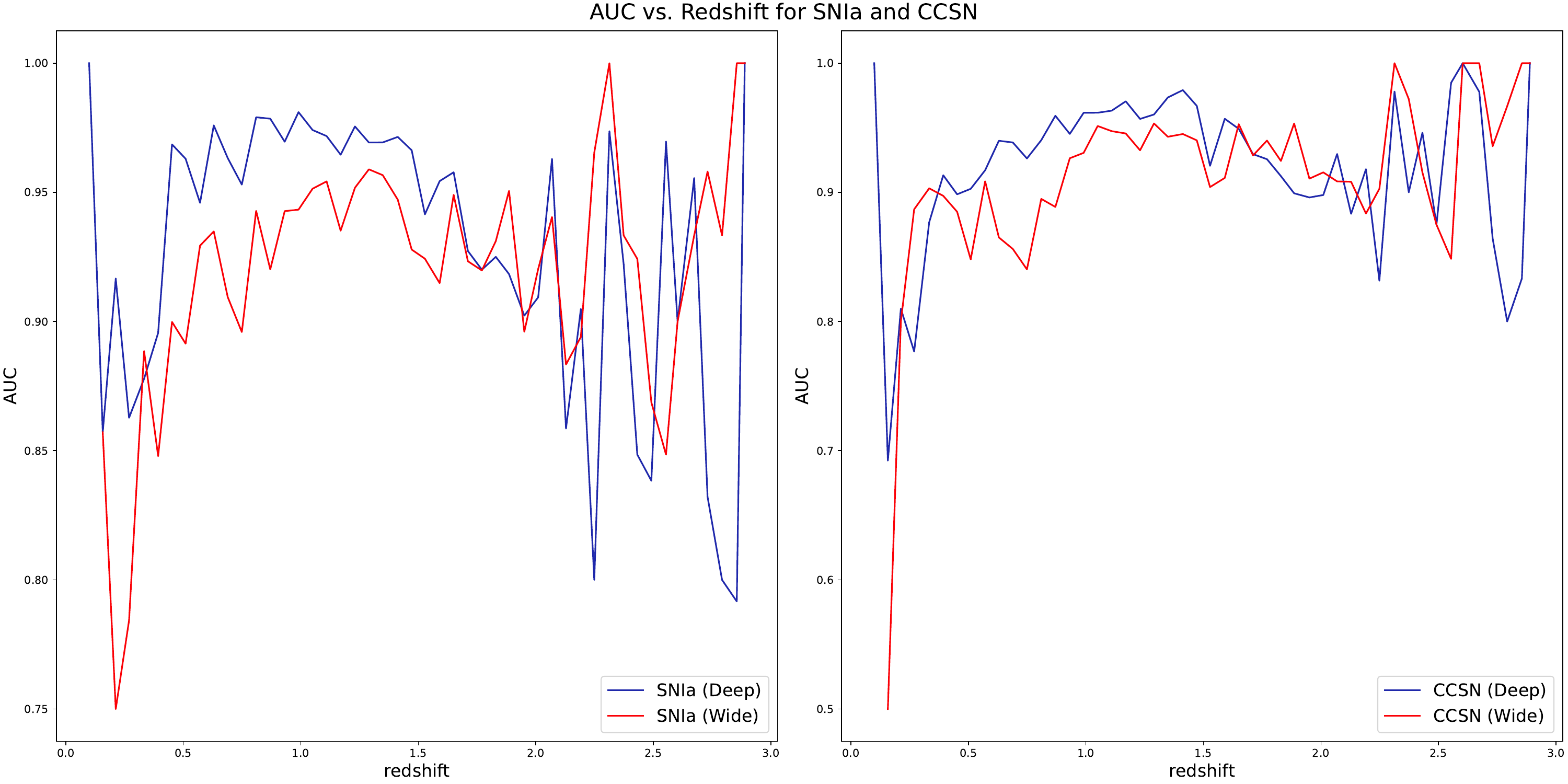}
    \caption{AUC vs. Redshift for SNIa (Left) and CCSN (Right) classifications using the \Parsnip classifier. Each panel shows AUC values across redshift ranges and two filter sets (Deep vs. Wide). AUC dips at low redshifts due to limited observations and declines beyond redshift 2.0 due to low representation. Deep filters outperform wide filters up to redshift $\sim 1.5$, after which performance converges.}
    \label{fig:AUCRedshift}
\end{figure*}

We gain deeper insights into classification performance by plotting AUC vs. redshift in \cref{fig:AUCRedshift}. This plot shows AUC values for SNe~Ia and CC~SNe across redshift ranges, highlighting performance variations with redshift. A key observation is the dip in AUC at lower redshifts, attributed to \rst's limited observations in these regions, resulting in less training data and reduced model accuracy. As redshift increases, AUC values generally improve, reflecting the model's enhanced ability to classify transient types with more available data. This trend emphasizes the importance of balanced datasets across all redshift ranges to ensure robust classifier performance.

We also compared the performance of the \Parsnip classifier across deep and wide observational survey tiers (see \cref{sec:dataset}). As shown in \cref{fig:AUCRedshift}, the classifier performed better in the deep tier up to a redshift of $z \approx 1.5$. Beyond $z \approx 2.0$, AUC drops significantly, likely due to low representation, as evident from the redshift distribution in \cref{fig:z_hist}. These observations suggest the need to prioritize low-redshift observations to improve model training and overall classification accuracy.

\FloatBarrier

\section{Discussion} \label{sec:disc}

\subsection{Key Objectives}
Our study highlights the effectiveness of the \Parsnip model in classifying transients based on photometric data from the \rst \romanplasticc dataset. The model's ability to generate latent representations that capture intrinsic features of transient types independently of redshift and observational conditions is particularly noteworthy. These findings align with previous research by \citep{Boone2021}, but our study extends this work by applying the model to the more recent \rst \romanplasticc dataset that also spans a larger redshift range. This is crucial for the \romanST mission of understanding dark energy through the observations of thousands of supernovae.

The visualization of latent parameters reveals distinct clusters for each transient type, demonstrating the model's ability to differentiate between them. This separation is particularly evident for well-represented types such as SN~Ia, CCSN, and 91bg-like supernovae. The performance of the \Parsnip model, as illustrated by confusion matrices and ROC curves, confirms its high accuracy in classifying these types. Despite the promising results, there are notable limitations. The dataset's imbalance, particularly the under-representation of certain transient types, poses challenges for classification accuracy. This limitation is evident in the lower performance metrics for less common types like SLSNe and TDEs.

The analysis of classifier performance across different redshift ranges shows a clear trend: while the model maintains high accuracy at higher redshifts, its performance diminishes at lower redshifts. This decline can be attributed to \rst's observational constraints at these ranges which limit the area to tens of square degrees, resulting in a smaller lower-redshift dataset. These findings emphasize the need for more comprehensive low-redshift observations to enhance model accuracy and reliability. Future work could focus on integrating additional datasets (such as ground-based optical data) and refining the model to address the challenges of underrepresented transient types. These suggestions align with strategies proposed by \citet{Alves2023} and \citet{Ishida2019}.

Overall, our study demonstrates the potential of machine learning models like \Parsnip in the field of astronomical data analysis. The ability to classify supernovae accurately is not only vital for the \romanST's objectives but also for broader astrophysical research. The insights gained from our analysis can guide future improvements in model design and observational strategies, ensuring that the \Parsnip model remains an effective tool for supernova classification.
\subsection{Limitations}
In discussing the classification performance of ParSNIP on the Roman dataset, it is essential to consider the model’s limitations, particularly regarding low-frequency transient classes like Pair-instability Supernovae (PISNe). The scarcity of these rare transients in the dataset affects ParSNIP’s ability to accurately learn and distinguish their unique characteristics. Consequently, ParSNIP may generalize the light curves of these rare classes to resemble more common classes, as seen with the model’s consistent misclassification of PISNe as Superluminous Supernovae (SLSNe). This tendency toward high-frequency classes highlights a critical limitation in the current dataset and poses a potential risk of systematic misclassifications in real-world applications \citep{Boone2021}.

Moreover, rare transients such as PISNe, Tidal Disruption Events (TDEs), and Superluminous Supernovae are known to exhibit considerable diversity in their light curves, which makes their classification inherently challenging. This diversity often results in broader variations within a class that are not well represented by a limited sample. For ParSNIP to learn and generalize these classes effectively, more comprehensive training data would be necessary to capture the full variability of their light curves. As a result, even if the model achieves seemingly good test performance on low-frequency classes, these results should be interpreted with caution, as they may not generalize well when applied to Roman’s live data stream.

These limitations have important implications for the deployment of ParSNIP in classifying transients from the Roman Space Telescope. A deployed model would likely struggle with accurately identifying these rare and diverse classes unless it is supplemented with additional data or other methods specifically tailored to address this scarcity. Stronger caveats should be applied to ParSNIP’s predictive power for these classes, cautioning that its classification performance is contingent upon data diversity and availability. This underlines the importance of either enriching the training dataset with more representative samples of low-frequency classes or exploring supplementary classification strategies to ensure robust and reliable performance across all transient types in the Roman survey.
\section{Conclusion} \label{sec:conc}

In this study, we employed the \citet{Boone2021} \Parsnip model and the \rst \romanplasticc dataset \citep{Rose2023} to assess the efficacy of machine learning in classifying transients for a possible \rst High Latitude Time Domain Survey (HLTDS) survey strategy. Using a variational autoencoder, \Parsnip effectively learned latent parameters that capture the intrinsic features of transient types, independent of redshift, resulting in high classification accuracy for well-represented transient types. \Parsnip's performance was confirmed through various metrics, including confusion matrices and ROC curves. Classification results over a multitude of redshifts and using separate filter types (deep vs wide) has shown that \Parsnip's classifier prefers the deep tier bands and redshifts lower than 1.5.

However, challenges persist in classifying underrepresented types and at lower redshifts, underscoring the importance of larger and more balanced datasets. While we have not simulated a cosmological analysis, the AUC of 0.973 for Type Ia supernovae is comparable to \plasticc results despite (spanning a much larger redshift range than \plasticc), so our findings suggest its feasibility. Furthermore, we assumed a perfectly representative training set, whereas real-world datasets tend to be biased toward lower-redshift, brighter events.

\Parsnip's ability to learn latent parameters invariant to redshift demonstrates its reliability in distinguishing between transient types across a wide range of redshifts. This capability highlights the potential for generalizing our methods to other large-scale photometric surveys, especially those with similar observational properties. Future work will explore the application of \Parsnip to additional datasets, including the original synthetic \plasticc dataset \citep{Kessler2019a}, which includes more lower-redshift transients, thereby providing a more uniform and representative redshift distribution. These results underscore the importance of balanced datasets in enhancing classification accuracy and the potential for ongoing refinement of the \Parsnip model. As we continue to optimize the model and incorporate more diverse datasets, \Parsnip is poised to become an increasingly valuable tool for \rst HLTDS and other astronomical surveys, contributing to a deeper understanding of the dynamic processes of the universe.

\begin{acknowledgments}
Funding for the Roman Supernova Project Infrastructure Team has been provided by NASA under contract to 80NSSC24M0023. We thank Ben Rose for early access to the \rst \romanplasticc simulation and feedback on this work.
\end{acknowledgments}

\software{
    Astropy \citep{astropy13, astropy18},
    Jupyter \citep{kluyver16},
    LightGBM \citep{ke17},
    Matplotlib \citep{hunter07},
    NumPy \citep{vanderwalt11},
    PyTorch \citep{pytorch},
    scikit-learn \citep{scikit-learn},
    SciPy \citep{scipy},
    SNCosmo \citep{barbary16c}
}

\bibliography{references}{}
\bibliographystyle{aasjournal}

\end{document}